\documentclass[prl,twocolumn,superscriptaddress,showkeys]{revtex4}
\usepackage{amssymb,epsf,epsfig}

\newcommand{\lsim}{
\mathrel{\hbox{\rlap{\hbox{\lower4pt\hbox{$\sim$}}}\hbox{$<$}}}}
\newcommand{\gsim}{
\mathrel{\hbox{\rlap{\hbox{\lower4pt\hbox{$\sim$}}}\hbox{$>$}}}}

\begin{document}
\begin{titlepage}
\vspace*{0.7truecm}
\begin{flushright}
CERN-PH-TH/2004-171\\
TUM-HEP-557/04\\
hep-ph/0409137
\end{flushright}

\vspace*{2.0truecm}
\begin{center}
\boldmath
{\Large\bf General Lower Bounds for $b\to d$ Penguin Processes}
\unboldmath
\end{center}

\vspace*{1.3truecm}

\begin{center}
{\bf \large Robert Fleischer${}^a$ and Stefan Recksiegel${}^b$}

\vspace{0.7truecm}

${}^a$ {\sl Theory Division, Department of Physics,
CERN, CH-1211 Geneva 23, Switzerland}

\vspace{0.2truecm}

${}^b$ {\sl Physik Department, Technische Universit\"at M\"unchen,
D-85748 Garching, Germany}

\end{center}

\vspace*{1.7cm}

\begin{center}
\large{\bf Abstract}

\vspace*{0.6truecm}

\begin{tabular}{p{14.5truecm}}
{\small 
For the exploration of flavour physics, $b\to d$ penguin processes are an 
important aspect, with the prominent example of $\bar B^0_d\to K^0\bar K^0$.
We recently derived lower bounds for the CP-averaged branching ratio of 
this channel in the Standard Model; they were found to be very close to the 
corresponding experimental upper limits, thereby suggesting that 
$\bar B^0_d\to K^0\bar K^0$ should soon be observed. In fact, the BaBar 
collaboration subsequently announced the first signals of this transition. 
Here we point out that it is also possible to derive lower bounds for 
$\bar B\to\rho\gamma$ decays, which are again surprisingly close to the 
current experimental upper limits. We show that these bounds are 
realizations of a general bound that holds within the Standard Model for 
$b\to d$ penguin processes, allowing further applications to decays of 
the kind $B^\pm\to K^{(\ast)\pm}K^{(\ast)}$ and 
$B^\pm\to\pi^\pm \ell^+ \ell^-$, $\rho^\pm \ell^+ \ell^-$.
}
\end{tabular}

\end{center}

\vspace*{3.7truecm}

\noindent
September 2004

\end{titlepage}

\newpage
\thispagestyle{empty}
\mbox{}

\newpage
\thispagestyle{empty}
\mbox{}

\rule{0cm}{23cm}

\newpage
\thispagestyle{empty}
\mbox{}

\setcounter{page}{0}

\preprint{hep-ph/0409137}
\preprint{CERN-PH-TH/2004-171}
\preprint{TUM-HEP-557/04}
\date{September 13, 2004}

\title{\boldmath General Lower Bounds for $b\to d$ 
Penguin Processes\unboldmath}

\author{Robert Fleischer}
\affiliation{Theory Division, Department of Physics,
CERN, CH-1211 Geneva 23, Switzerland}

\author{Stefan Recksiegel}
\affiliation{Physik Department, Technische Universit\"at M\"unchen,
D-85748 Garching, Germany}

\begin{abstract}
\vspace{0.2cm}\noindent
For the exploration of flavour physics, $b\to d$ penguin processes are an 
important aspect, with the prominent example of $\bar B^0_d\to K^0\bar K^0$.
We recently derived lower bounds for the CP-averaged branching ratio of 
this channel in the Standard Model; they were found to be very close to the 
corresponding experimental upper limits, thereby suggesting that 
$\bar B^0_d\to K^0\bar K^0$ should soon be observed. In fact, the BaBar 
collaboration subsequently announced the first signals of this transition. 
Here we point out that it is also possible to derive lower bounds for 
$\bar B\to\rho\gamma$ decays, which are again surprisingly close to the 
current experimental upper limits. We show that these bounds are 
realizations of a general bound that holds within the Standard Model for 
$b\to d$ penguin processes, allowing further applications to decays of 
the kind $B^\pm\to K^{(\ast)\pm}K^{(\ast)}$ and 
$B^\pm\to\pi^\pm \ell^+ \ell^-$, $\rho^\pm \ell^+ \ell^-$.
\end{abstract}


\maketitle

Despite the tremendous progress at the $B$ factories, we still have 
few insights into the rare decays that are mediated by $b\to d$ penguin 
topologies, which represent a key element in the testing of the
quark-flavour sector of the Standard Model (SM) that is described by the
Cabibbo--Kobayashi--Maskawa (CKM) matrix \cite{CKM}. An important 
example is the decay $\bar B^0_d\to K^0\bar K^0$, which originates from 
$b\to d \bar ss$ flavour-changing neutral-current (FCNC) processes. Within 
the SM, these are governed by QCD penguin-like topologies, so that we may write
\begin{equation}\label{Ampl-BdK0K0}
A(\bar B^0_d\to K^0\bar K^0)\propto
{\cal P}_{tc}^{K\!K}\left[1-\rho_{K\!K} e^{i\theta_{K\!K}}e^{-i\gamma}\right],
\end{equation}
where $\gamma$ is the usual angle of the unitarity triangle of the CKM 
matrix, whereas ${\cal P}_{tc}^{K\!K}$ and $\rho_{K\!K}e^{i\theta_{K\!K}}$ 
are CP-conserving hadronic parameters. The latter quantities not only 
depend on the penguin topologies with internal top-quark exchanges, but
are also expected to be affected significantly by those with internal up- 
and charm-quark exchanges \cite{RF-BdKK}, containing also long-distance
rescattering effects \cite{BFM}. Consequently, it seems essentially 
impossible to calculate ${\cal P}_{tc}^{K\!K}$ and 
$\rho_{K\!K}e^{i\theta_{K\!K}}$ in a reliable manner from first principles,
despite theoretical progress \cite{Be-Ne}.

In a recent paper \cite{FR}, we addressed this problem, and pointed out 
that {\it lower} bounds for the CP-averaged branching ratio 
BR$(B_d\to K^0\bar K^0)$ can be derived in the SM. To this end, we assume 
that $\gamma=(65\pm7)^\circ$, as in the SM \cite{CKM-Book}, consider 
$\rho_{K\!K}$ and $\theta_{K\!K}$ as ``unknown'' parameters, i.e.\ vary 
them within their whole physical range, and fix $|{\cal P}_{tc}^{K\!K}|$ 
with the help of the $SU(3)$ flavour symmetry of strong interactions. 
The strategy developed in \cite{BFRS} offers the following two avenues, 
using data for
\begin{itemize}
\item[i)] $B\to\pi\pi$ ($b\to d$) decays: 
\begin{equation}\label{BdKK-bound1}
\mbox{BR}(B_d\to K^0\bar K^0)_{\rm min}= 
\Xi^K_\pi\times\left(1.39\,^{+1.54}_{-0.95}\right) \times 10^{-6},
\end{equation}
\item[ii)] $B\to\pi K$ ($b\to s$) decays, 
complemented by the $B\to\pi\pi$ system
to determine a small correction:
\begin{equation}\label{BdKK-bound2}
\mbox{BR}(B_d\to K^0\bar K^0)_{\rm min}= 
\Xi^K_\pi\times\left(1.36\,^{+0.18}_{-0.21}\right) \times 10^{-6}.
\end{equation}
\end{itemize}
Here we have included factorizable $SU(3)$-breaking corrections, making 
their impact explicit through
\begin{equation}\label{Xi-K-pi}
\Xi^K_\pi=\left[\frac{f_0^K}{0.331}\frac{0.258}{f_0^\pi}\right]^2;
\end{equation}
the numerical values for the $B\to K,\pi$ form factors $f_0^{K,\pi}$ 
refer to a recent light-cone sum-rule analysis \cite{Ball}. Comparing 
(\ref{BdKK-bound1}) and (\ref{BdKK-bound2}) with the experimental 
{\it upper} bound of $\mbox{BR}(B_d\to K^0\bar K^0)<1.5\times 10^{-6}$ 
($90\%$ C.L.), we concluded that $\bar B^0_d\to K^0\bar K^0$ should soon 
be observed. 

In fact, the BaBar collaboration subsequently reported 
the first signals, with the CP-averaged branching ratio 
$\mbox{BR}(B_d\to K^0\bar K^0)=(1.19^{\,+0.40}_{\,-0.35} \pm 0.13) 
\times 10^{-6}$ \cite{BaBar-BKK}. This is a very exciting measurement, 
as it establishes -- for the first time -- a $b\to d$ penguin process.
The consistency 
of the BaBar result \cite{BaBar-BKK} with (\ref{BdKK-bound1}) and 
(\ref{BdKK-bound2}) is a first successful test of the SM description of the 
$b\to d \bar ss$ FCNC processes, although the current uncertainties are  
still large; using the most recent data \cite{ICHEP04,HFAG}, the 
numerical factors in (\ref{BdKK-bound1}) and (\ref{BdKK-bound2})
are modified as $(1.15\,^{+1.13}_{-0.77}) \times 10^{-6}$ and 
$(1.37\,^{+0.16}_{-0.20}) \times 10^{-6}$, respectively. More powerful tests 
will be possible in the future, where also the CP-violating 
$B_d\to K^0\bar K^0$ asymmetries will play a key r\^ole \cite{FR}; a 
specific new-physics analysis within the framework of supersymmetry was 
very recently performed in \cite{GiMo}.

Another important tool to explore the $b\to d$ penguin sector is
provided by $\bar B\to\rho\gamma$ modes (for recent analyses, see 
\cite{ALP-rare,BoBu}). The current experimental picture of their 
CP-averaged branching ratios is given as follows:
\begin{equation}\label{Brho-gam-char-EXP}
\mbox{BR}(B^\pm\to \rho^\pm\gamma)<\left\{\begin{array}{ll}
1.8\times 10^{-6} & \mbox{(BaBar \cite{Babar-Brhogamma})}\\
2.2\times 10^{-6} & \mbox{(Belle \cite{Belle-Brhogamma})}\\
\end{array} \right.
\end{equation}
\begin{equation}\label{Brho-gam-neut-EXP}
\mbox{BR}(B_d\to \rho^0\gamma)<\left\{\begin{array}{ll}
0.4\times 10^{-6} & \mbox{(BaBar \cite{Babar-Brhogamma})}\\
0.8\times 10^{-6} & \mbox{(Belle \cite{Belle-Brhogamma}),}\\
\end{array} \right.
\end{equation}
where the upper bounds are at the $90\%$ confidence level. In 
the SM, these decays are described by a low-energy effective Hamiltonian
of the following structure \cite{BBL-rev}:
\begin{equation}\label{Ham-bdgam}
{\cal H}_{\rm eff}^{b\to d\gamma}=\frac{G_{\rm F}}{\sqrt{2}}
\sum_{j=u,c} \! V_{jd}^\ast V_{jb}\left[\sum_{k=1}^{2}C_k Q_k^{jd}\!+\!
\sum_{k=3}^{8}C_k Q_k^{d}\right].
\end{equation}
Here the $Q_{1,2}^{jd}$ denote the current--current operators, whereas the 
$Q_{3\ldots 6}^{d}$ are the QCD penguin operators, which govern the
decay $\bar B^0_d\to  K^0\bar K^0$ together with the
penguin-like contractions of $Q_{1,2}^{cd}$ and $Q_{1,2}^{ud}$. On the
other hand,
\begin{equation}
Q_{7,8}^{d}=\frac{1}{8\pi^2}m_b\bar d_i \sigma^{\mu\nu}(1+\gamma_5)
\left\{e b_i F_{\mu\nu} ,\, g_{\rm s}T^a_{ij}b_j G^a_{\mu\nu} \right\}
\end{equation}
are the electro- and chromomagnetic penguin operators. 

Using (\ref{Ham-bdgam}) and the Wolfenstein
parametrization of the CKM matrix \cite{wolf}, generalized to the
next-to-leading order in $\lambda=0.224$ \cite{BLO}, we may write
\begin{equation}\label{Ampl-Brhogam}
A(\bar B \to \rho\gamma)=c_\rho \lambda^3 A {\cal P}_{tc}^{\rho\gamma}
\left[1-\rho_{\rho\gamma}e^{i\theta_{\rho\gamma}}e^{-i\gamma}\right],
\end{equation}
where $c_\rho=1/\sqrt{2}$ and 1 for $\rho=\rho^0$ and $\rho^\pm$,
respectively, $A=|V_{cb}|/\lambda^2$, ${\cal P}_{tc}^{\rho\gamma}\equiv
{\cal P}_t^{\rho\gamma}-{\cal P}_c^{\rho\gamma}$, and
\begin{equation}
\rho_{\rho\gamma}e^{i\theta_{\rho\gamma}}\equiv R_b\left[
\frac{{\cal P}_t^{\rho\gamma}-
{\cal P}_u^{\rho\gamma}}{{\cal P}_t^{\rho\gamma}-
{\cal P}_c^{\rho\gamma}}\right].
\end{equation}
The ${\cal P}_j^{\rho\gamma}$ are strong amplitudes, which 
have the following interpretation: ${\cal P}_u^{\rho\gamma}$ and
${\cal P}_c^{\rho\gamma}$ refer to the matrix elements of
$\sum_{k=1}^{2}C_k Q_k^{ud}$ and $\sum_{k=1}^{2}C_k Q_k^{cd}$, 
respectively, whereas ${\cal P}_t^{\rho\gamma}$ corresponds to 
$-\sum_{k=3}^{8}C_k Q_k^{d}$. Consequently, ${\cal P}_u^{\rho\gamma}$
and ${\cal P}_c^{\rho\gamma}$ describe the penguin topologies with
internal up- and charm-quark exchanges, respectively, whereas 
${\cal P}_t^{\rho\gamma}$ corresponds to the penguins with the top
quark running in the loop. Finally, $R_b\propto |V_{ub}/V_{cb}|$ is
one side of the unitarity triangle \cite{BLO}. Let us note that 
(\ref{Ampl-Brhogam}) refers to a given photon helicity. However, 
the $b$ quarks couple predominantly to left-handed photons in 
the SM, so that the right-handed amplitude is usually neglected \cite{GP}; 
we shall return to this point below. Comparing (\ref{Ampl-Brhogam}) with 
(\ref{Ampl-BdK0K0}), we observe that the structure of both amplitudes is 
the same. In analogy to $\rho_{K\!K} e^{i\theta_{K\!K}}$, 
$\rho_{\rho\gamma}e^{i\theta_{\rho\gamma}}$ may also be affected by 
long-distance effects, which represent a key uncertainty of 
$\bar B\to\rho\gamma$ decays \cite{GP,LHC-Book}. 

If we replace all down quarks in (\ref{Ham-bdgam}) by strange 
quarks, we obtain the effective Hamiltonian for $b\to s\gamma$ 
processes, which are already well established 
experimentally \cite{HFAG}:
\begin{eqnarray}
\mbox{BR}(B^\pm\to K^{\ast\pm}\gamma)&=&(40.3\pm2.6)\times 
10^{-6}\label{BR-charged}\\
\mbox{BR}(B_d^0\to K^{\ast0}\gamma)&=&(40.1\pm2.0)\times 
10^{-6}.\label{BR-neutral}
\end{eqnarray}
In analogy to (\ref{Ampl-Brhogam}), we may write
\begin{equation}\label{Ampl-BKastgam}
A(\bar B \!\to\! K^\ast \!\gamma)\!=-\!
\frac{\lambda^3 \! A {\cal P}_{tc}^{K^\ast\!\gamma}}{\sqrt{\epsilon}} \!
\left[1\!+\!\epsilon\rho_{K\!^\ast\!\gamma}e^{i\theta_{K\!^\ast\!\gamma}}
e^{-i\!\gamma}\right]\!,
\end{equation}
with $\epsilon\equiv\lambda^2/(1-\lambda^2)=0.053$. Thanks to the smallness
of $\epsilon$, the parameter 
$\rho_{K\!^\ast\gamma}e^{i\theta_{K\!^\ast\gamma}}$ 
plays an essentially negligible r\^ole for the $\bar B \to K^\ast \gamma$ 
transitions.

Let us first focus on the charged decays $B^\pm \to \rho^{\pm} \gamma$ 
and $B^\pm \to K^{\ast\pm} \gamma$. Here we obtain
\begin{equation}\label{rare-ratio}
\frac{\mbox{BR}(B^\pm \to \rho^{\pm} 
\gamma)}{\mbox{BR}(B^\pm \to K^{\ast\pm} \gamma)}=\epsilon
\left[\frac{\Phi_{\rho\gamma}}{\Phi_{K\!^\ast\gamma}}\right]
\left|\frac{{\cal P}_{tc}^{\rho\gamma}}{{\cal P}_{tc}^{K\!^\ast\gamma}}
\right|^2 H^{\rho\gamma}_{K\!^\ast\gamma},
\end{equation}
where $\Phi_{\rho\gamma}$ and $\Phi_{K\!^\ast\gamma}$ denote phase-space 
factors, and 
\begin{equation}
H^{\rho\gamma}_{K\!^\ast\gamma}\equiv
\frac{1-2\rho_{\rho\gamma}\cos\theta_{\rho\gamma}\cos\gamma+
\rho_{\rho\gamma}^2}{1+2\epsilon\rho_{K\!^\ast\gamma}
\cos\theta_{K\!^\ast\gamma}
\cos\gamma+\epsilon^2\rho_{K\!^\ast\gamma}^2}.
\end{equation}
If we apply now the $U$-spin flavour symmetry of strong interactions to 
these rare decays \cite{HM}, which is a subgroup of flavour $SU(3)$ that 
relates down and strange quarks in the same manner as the conventional 
isospin symmetry relates down and up quarks, we obtain 
\begin{equation}\label{U-spin1}
|{\cal P}_{tc}^{\rho\gamma}|=|{\cal P}_{tc}^{K\!^\ast\gamma}|
\end{equation}
\begin{equation}\label{U-spin2}
\rho_{\rho\gamma}e^{i\theta_{\rho\gamma}}=
\rho_{K\!^\ast\gamma}e^{i\theta_{K\!^\ast\gamma}}\equiv
\rho e^{i\theta}.
\end{equation}
Although (\ref{U-spin1}) allows us to determine the ratio of the
penguin amplitudes $|{\cal P}_{tc}|$ in (\ref{rare-ratio}) -- up to 
$SU(3)$-breaking effects to be discussed below -- we are still left
with the dependence on $\rho$ and $\theta$. However, if we keep $\rho$ 
and $\theta$ as free parameters, we may show that 
$H^{\rho\gamma}_{K\!^\ast\gamma}$ satisfies
\begin{equation}\label{H-bound}
H^{\rho\gamma}_{K\!^\ast\gamma}\geq \left[1-2\epsilon
\cos^2\gamma+{\cal O}(\epsilon^2)\right]\sin^2\gamma,
\end{equation}
where the term linear in $\epsilon$ gives a shift of about $1.9\%$. 

Concerning possible $SU(3)$-breaking effects to (\ref{U-spin2}), they 
may only enter this tiny correction and are negligible for our analysis. 
On the other hand, the $SU(3)$-breaking corrections to (\ref{U-spin1}) 
have a sizeable impact. Following \cite{ALP-rare,BoBu}, we write
\begin{equation}\label{SU3-break-rare}
\left[\frac{\Phi_{\rho\gamma}}{\Phi_{K\!^\ast\gamma}}\right]
\left|\frac{{\cal P}_{tc}^{\rho\gamma}}{{\cal P}_{tc}^{K\!^\ast\gamma}}
\right|^2=\left[\frac{M_B^2-M_\rho^2}{M_B^2-M_{K^\ast}^2}\right]^3
\zeta^2,
\end{equation}
where $\zeta=F_\rho/F_{K^\ast}$ is the $SU(3)$-breaking ratio of the
$B^\pm\to\rho^\pm\gamma$ and $B^\pm\to K^{\ast\pm}\gamma$ form factors; a 
light-cone sum-rule analysis gives $\zeta^{-1}=1.31\pm0.13$ \cite{Ball-Braun}, 
to be compared with the result $\zeta^{-1}=1.1\pm0.1$ of a preliminary lattice 
analysis \cite{becirevic}. Consequently, (\ref{H-bound}) and 
(\ref{SU3-break-rare}) allow us to convert the measured 
$B^\pm\to K^{\ast\pm}\gamma$ branching ratio (\ref{BR-charged}) 
into a {\it lower} SM bound for 
$\mbox{BR}(B^\pm\to\rho^\pm\gamma)$ with the help of (\ref{rare-ratio}). 
If we use the SM range $\gamma=(65\pm7)^\circ$ 
\cite{CKM-Book} and $\zeta^{-1}=1.31\pm0.13$ \cite{Ball-Braun}, we obtain
\begin{equation}\label{Brhogam-char}
\mbox{BR}(B^\pm\to \rho^\pm\gamma)_{\rm min}=\left(1.02\,^{+0.27}_{-0.23}
\right)\times10^{-6}.
\end{equation}

A similar kind of reasoning holds also for the $U$-spin pairs 
$B^\pm\to K^\pm K, \pi^\pm K$ and $B^\pm\to K^\pm K^\ast, \pi^\pm K^\ast$.
In the former case, the factorizable $SU(3)$-breaking effects are governed 
by (\ref{Xi-K-pi}). In the latter case, following \cite{Ball}, the form 
factors $f_+^{K,\pi}$ enter. However, because of $f_+^P=f_0^P$, 
we arrive again at (\ref{Xi-K-pi}). Using then the experimental results 
$\mbox{BR}(B^\pm\to \pi^\pm K)=(24.1\pm1.3)\times10^{-6}$ and 
$\mbox{BR}(B^\pm\to \pi^\pm K^\ast)=(9.76\,^{+1.16}_{-1.22})\times10^{-6}$
\cite{HFAG}, we obtain
\begin{eqnarray}
\mbox{BR}(B^\pm\!\to\! K^\pm K)_{\rm min} \!\!&=&\!\! \Xi^K_\pi\!\times\!
\left(1.69\,^{+0.21}_{-0.24}\right)\!\times\! 10^{-6}\label{BKK-char}\\
\mbox{BR}(B^\pm\!\to\! K^\pm K^\ast)_{\rm min} \!\!&=&\!\! \Xi^K_\pi\!\times\!
\left(0.68\,^{+0.11}_{-0.13}
\right)\!\times \! 10^{-6}.\label{BpiKast}
\end{eqnarray}
In the case of $B^\pm\to K^\pm K$, the lower SM bound is very close to the 
experimental upper bound of $2.4\times 10^{-6}$ \cite{BaBar-BKK},
whereas the upper limit of $5.3\times 10^{-6}$ for $B^\pm\to K^\pm K^\ast$
still leaves a lot of space. Obviously, we may also consider the
$B^\pm\to K^{\ast\pm} K, \rho^\pm K$ system. Using the $B\to V$ form factors
obtained in \cite{Ball-Braun} to deal with the factorizable $SU(3)$-breaking 
effects, we arrive at 
\begin{equation}\label{BrhoK}
\frac{\mbox{BR}(B^\pm\to K^{\ast\pm} K)}{\mbox{BR}(B^\pm\to \rho^\pm K)}
\geq \Xi^{K^\ast}_\rho \! \times 0.084 \times 
\left[1-2\epsilon\cos^2\gamma\right]
\sin^2\gamma,
\end{equation}
with
\begin{equation}\label{Xi-Kast-rho}
\Xi^{K^\ast}_\rho=\left[\frac{A_0^{K^\ast}}{0.470}
\frac{0.372}{A_0^\rho}\right]^2.
\end{equation}
Although the individual form factors are very different, (\ref{Xi-Kast-rho}) 
yields essentially the same correction as (\ref{Xi-K-pi}). Since only the 
upper bound of $\mbox{BR}(B^\pm\to \rho^\pm K)<48\times 10^{-6}$ is 
available, we may not yet apply (\ref{BrhoK}). 

Let us now turn to $\bar B^0_d\to\rho^0\gamma$, which receives 
contributions from exchange and penguin annihilation topologies that are 
not present in 
$\bar B^0_d\to \bar K^{\ast0}\gamma$; in the case of $B^\pm\to\rho^\pm\gamma$ 
and $B^\pm\to K^{\ast\pm}\gamma$, which are related by the $U$-spin symmetry, 
there is a one-to-one correspondence of topologies \cite{GP}. Making the 
plausible assumption that the topologies involving the spectator quarks play 
a minor r\^ole, and taking the factor of $c_{\rho^0}=1/\sqrt{2}$ in 
(\ref{Ampl-Brhogam}) into account, the counterpart of (\ref{Brhogam-char}) 
is given by 
\begin{equation}\label{Brhogam-neut}
\mbox{BR}(B_d\to \rho^0\gamma)_{\rm min}=\left(0.51\,^{+0.13}_{-0.11}
\right)\times10^{-6}.
\end{equation}

If we compare the {\it lower} SM bounds in (\ref{Brhogam-char}) and
(\ref{Brhogam-neut}) with the current experimental {\it upper} bounds in
(\ref{Brho-gam-char-EXP}) and (\ref{Brho-gam-neut-EXP}), respectively, 
we observe that they are remarkably close to one another. Consequently,
we expect that the $\bar B\to\rho\gamma$ modes should soon be discovered 
at the $B$ factories. The next important step would then be the measurement
of their CP-violating observables.

The authors of \cite{ALP-rare,BoBu} followed a different avenue to 
confront the experimental bounds in 
(\ref{Brho-gam-char-EXP}) and (\ref{Brho-gam-neut-EXP}) with the SM,
converting them into upper bounds for the side $R_t\propto |V_{td}/V_{cb}|$
of the unitarity triangle \cite{BLO}. To this end, they use also 
(\ref{SU3-break-rare}),
and calculate the CP-conserving (complex) parameter $\delta a$ entering
$\rho_{\rho\gamma}e^{i\theta_{\rho\gamma}}=R_b\left[1+\delta a\right]$
with the help of QCD factorization. The corresponding result, which
favours a small impact of $\delta a$, takes leading and next-to-leading 
order QCD corrections into account and holds to leading order in the 
heavy-quark limit \cite{BoBu}. However, in view of the remarks about possible 
long-distance effects made above and the $B$-factory data for the $B\to\pi\pi$
system, which indicate large corrections to the QCD factorization picture
for non-leptonic $B$ decays into two light pseudoscalar mesons 
\cite{BFRS,Bpipi-non-fact}, it is not obvious to us that the impact of 
$\delta a$ is actually small. The advantage of our bound following from 
(\ref{H-bound}) is that it is  -- by construction -- {\it not} affected by 
$\rho_{\rho\gamma}e^{i\theta_{\rho\gamma}}$ at all.

Interestingly, the lower bounds for the CP-averaged 
$B^\pm\to K^{(\ast)\pm}K^{(\ast)}$ and $B\to\rho\gamma$ branching 
ratios discussed above are actually 
realizations of a general bound that can be derived
in the SM for $b\to d$ penguin processes. Let us consider such a decay,
$\bar B \to \bar f_d$. In analogy to (\ref{Ampl-BdK0K0}) and 
(\ref{Ampl-Brhogam}), we may then write
\begin{equation}
A(\bar B \to \bar f_d)= A^{(0)}_d
\left[1-\rho_de^{i\theta_d}e^{-i\gamma}\right],
\end{equation}
so that the CP-averaged amplitude square takes the form
\begin{equation}
\langle|A(B \to f_d)|^2\rangle=|A^{(0)}_d|^2
\left[1-2\rho_d\cos\theta_d\cos\gamma+\rho_d^2\right].
\end{equation}
In general, $\rho_d$ and $\theta_d$ depend on the point in phase space
considered. This has the implication that the expression 
\begin{equation}
\mbox{BR}(B \to f_d)=\tau_B\left[\sum_{\rm Pol}
\int \!\! d \, {\rm PS} \, \langle|A(B \to f_d)|^2\rangle \right]
\end{equation}
for the CP-averaged branching ratio, where the sum runs over possible
polarization configurations of $f_d$, does {\it not} factorize into 
$|A^{(0)}_d|^2$ and $[1-2\rho_d\cos\theta_d\cos\gamma+\rho_d^2]$ as 
in the case of the two-body decays considered above. However, if we 
keep $\rho_d$ and $\theta_d$ as free, ``unknown'' parameters at any 
given point in phase space, we obtain
\begin{equation}
\langle|A(B \to f_d)|^2\rangle\geq|A^{(0)}_d|^2 \sin^2\gamma,
\end{equation}
which implies
\begin{equation}
\mbox{BR}(B \to f_d)\geq\tau_B\left[\sum_{\rm Pol}
\int \!\! d \, {\rm PS} \, |A^{(0)}_d|^2 \right]\sin^2\gamma.
\end{equation}

We consider now a $b\to s$ penguin process $\bar B \to \bar f_s$, 
which is the counterpart of $\bar B \to \bar f_d$ in 
that the corresponding CP-conserving strong amplitudes can be related
to one another through the $SU(3)$ flavour symmetry. In analogy to 
(\ref{Ampl-BKastgam}), we then have
\begin{equation}
A(\bar B \to \bar f_s)= - \frac{A^{(0)}_s}{\sqrt{\epsilon}}
\left[1+\epsilon\rho_s e^{i\theta_s}e^{-i\gamma}\right].
\end{equation}
If we neglect the term proportional to $\epsilon$ in the square bracket, 
we arrive at
\begin{equation}\label{general-bound}
\frac{\mbox{BR}(B \to f_d)}{\mbox{BR}(B \to f_s)}
\geq \epsilon \left[\frac{\sum_{\rm Pol}\int \! d \, {\rm PS} \, 
|A^{(0)}_d|^2 }{\sum_{\rm Pol}\int \! d \, {\rm PS} \, |A^{(0)}_s|^2 }
\right]\sin^2\gamma.
\end{equation}
Apart from the tiny $\epsilon$ correction, which gave a shift of about
$1.9\%$ in (\ref{H-bound}), (\ref{general-bound}) is valid
exactly in the SM. If we now apply the $SU(3)$ flavour symmetry, we obtain
\begin{equation}\label{SU3-limit}
\frac{\sum_{\rm Pol}\int \! d \, {\rm PS} \, 
|A^{(0)}_d|^2 }{\sum_{\rm Pol}\int \! d \, {\rm PS} \, |A^{(0)}_s|^2 }
\stackrel{SU(3)_{\rm F}}{\longrightarrow} 1.
\end{equation}
Since, in the SM, $\sin^2\gamma\sim0.8$ is favourably large and the decay
$\bar B \to \bar f_s$ will be measured before its $b\to d$ 
counterpart  -- simply because of the CKM enhancement -- 
(\ref{general-bound}) provides strong lower bounds for 
$\mbox{BR}(B \to f_d)$. 

It is instructive to return briefly to $B\to\rho\gamma$. Looking at 
(\ref{general-bound}), we observe immediately that the assumption that 
these modes are governed by a single photon helicity is no longer 
required. Consequently, (\ref{Brhogam-char}) and (\ref{Brhogam-neut}) 
are actually very robust with respect to this issue, which may only affect 
the $SU(3)$-breaking corrections to a small extend. 

We may now also complement the bounds in (\ref{BKK-char})--(\ref{BrhoK}) 
through the $B^\pm\to K^{\ast\pm}K^{\ast}, \rho^\pm K^\ast$ system, where 
we have to sum in (\ref{general-bound}) over three polarization configurations
of the vector mesons. The analysis of the $SU(3)$-breaking corrections is
now more involved. However, if we expand in $M_\rho/M_B$ and $M_{K^\ast}/M_B$, 
and use the form-factor relation $A_3^V=A_0^V$ \cite{Ball-Braun}, we find that 
the factorizable corrections are described to a good approximation by 
(\ref{Xi-Kast-rho}). Using then the very recent result of 
$\mbox{BR}(B^\pm\to \rho^\pm K^\ast)=(9.2\pm2.0)\times10^{-6}$ \cite{HFAG},
we obtain
\begin{equation}\label{B-Kast-rho}
\mbox{BR}(B^\pm\to K^{\ast\pm} K^\ast)_{\rm min}\approx
\Xi^{K^\ast}_\rho \!\! \times \left(0.64^{+0.15}_{-0.16}\right)\times10^{-6};
\end{equation}
the current experimental upper bound reads $71\times 10^{-6}$. 
Interestingly, (\ref{B-Kast-rho}) would be reduced by $\sim 0.6$ in 
the strict $SU(3)$ limit, i.e.\ would be more conservative. A similar 
comment applies to (\ref{BdKK-bound1}), (\ref{BdKK-bound2}) and  
(\ref{BKK-char})--(\ref{BrhoK}). On the other hand, the 
$B\to\rho\gamma$ bounds in (\ref{Brhogam-char}) and 
(\ref{Brhogam-neut}) would be enhanced by $\sim 1.7$ in this case.
However, here the theoretical situation is more favourable since we 
have not to rely on the factorization hypothesis to deal with the 
$SU(3)$-breaking effects as in the non-leptonic decays \cite{FR}. 

Another interesting application of (\ref{general-bound}) is offered by 
decays of the kind $\bar B\to \pi \ell^+\ell^-$ and 
$\bar B\to \rho \ell^+\ell^-$. It is
well known that the $\rho_d$ terms complicate the interpretation of
the corresponding data considerably \cite{LHC-Book}; our bound offers
SM tests that are not affected by these contributions. The 
structure of the $b\to d \ell^+\ell^-$ Hamiltonian is similar to 
(\ref{Ham-bdgam}), but involves the additional operators
\begin{equation}
Q_{9,10}=\frac{\alpha}{2\pi}(\bar\ell\ell)_{\rm V\!,\,A}
(\bar d_i b_i)_{\rm V-A}.
\end{equation}
The $b \to s$ counterparts of these transitions, $\bar B\to K \ell^+\ell^-$ 
and $\bar B\to K^\ast \ell^+\ell^-$, were already observed at the $B$ 
factories, with branching ratios at the $0.6\times 10^{-6}$ and 
$1.4\times 10^{-6}$ levels \cite{HFAG}, respectively, and received a lot of 
theoretical attention (see, for instance, \cite{BKll}). For the application
of (\ref{general-bound}), the charged decay combinations
$B^\pm\to \pi^\pm \ell^+\ell^-, K^\pm \ell^+\ell^-$ and
$B^\pm\to \rho^\pm \ell^+\ell^-, K^{\ast\pm} \ell^+\ell^-$ are suited
best since the corresponding decay pairs are related to each other 
through the $U$-spin symmetry \cite{HM}. We strongly encourage 
detailed studies of the associated $SU(3)$-breaking corrections to 
(\ref{SU3-limit}) and are confident that we will have a good picture
of these effects by the time the $B^\pm\to \pi^\pm \ell^+\ell^-$, 
$\rho^\pm \ell^+\ell^-$ modes will come within experimental reach.

Should the $b\to d$ penguin decays actually be found in accordance 
with the bounds derived above, as in the case of $\bar B^0_d\to K^0\bar K^0$,
we would have a first confirmation of the SM description of the 
corresponding FCNC processes. 
On the other hand, it would be much more exciting if some bounds 
should be significantly violated through the destructive interference 
between NP and SM contributions. As the various decay classes are 
governed by different operators, we may well encounter surprises.

\vspace*{1mm}
{\bf Acknowledgements}
This work was supported in part by the German Bundesministerium 
f\"ur Bildung und Forschung under the contract 05HT4WOA/3 and the 
DFG project Bu.\ 706/1-2. S.R. is grateful to CERN-PH-TH for 
hospitality while this paper was finalized.

\end{document}